\documentclass[twocolumn,pra,showpacs,preprintnumbers,amsmath,amssymb,showkeys,superscriptaddress]{revtex4-1}

\usepackage{dcolumn}
\usepackage{bezier}
\usepackage[title]{appendix}
\usepackage{graphicx,amsmath,amssymb,bm}

\newcommand{\bra}[1]{\left<{#1}\right|}
\newcommand{\ket}[1]{\left|{#1}\right>}
\newcommand{\braket}[2]{\left<\left.{#1}\right|{#2}\right>}

\usepackage[usenames]{color}
\usepackage{soul}
\setulcolor{red}
\setstcolor{red}
\sethlcolor{yellow}

\begin{document}

\title{Null weak values in multi-level systems} 
\author{Oded Zilberberg}
\affiliation{Department of Condensed Matter Physics, Weizamnn Institute of Science, Rehovot 76100, Israel}
\author{Alessandro Romito}
\affiliation{\mbox{Dahlem Center for Complex Quantum Systems and Fachbereich Physik, Freie Universit\"at Berlin, 14195 Berlin, Germany}}
\author{Yuval Gefen}
\affiliation{Department of Condensed Matter Physics, Weizamnn Institute of Science, Rehovot 76100, Israel}

\date{\today}


\begin{abstract}
A two-step measurement protocol of a quantum system, known as \textit{weak value} (WV), has been introduced more than two decades ago by Aharonov \textit{et al}.~\cite{Aharonov:1988aa}, and has since been studied in various contexts. Here we discuss another two-step measurement protocol which we dub \textit{null weak value} (NWV). The protocol consists of a partial-collapse measurement followed by quantum manipulation on the system and finally a strong measurement. The first step is a strong measurement which takes place with small probability. The second strong measurement is used as postselection on the outcome of the earlier step. Not being
measured in the partial-collapse stage (null outcome) leads to a non-trivial correlation between the two measurements. The NVW protocol, first defined for a two-level system~\cite{Zilberberg:12}, is generalized to a multi-level system, and compared to the standard-WV protocol. 
\end{abstract}


\maketitle

\section{Introduction}

A measurement in quantum mechanics is a probabilistic
process which, in the simplest situation, is described by the
projection postulate~\cite{von-neumann}.
Conditional quantum measurements, however, can lead to results that
cannot be interpreted in terms of classical probabilities, due to the
quantum correlations between measurements.
An intriguing example for correlated quantum measurements outcome is the so called \emph{weak
  values} (WVs). It is the outcome of a measurement scheme originally
developed by Aharonov, Albert and Vaidman~\cite{Aharonov:1988aa}.
The weak value measurement protocol consists of (i) initializing the system in
a certain state  $|\, i\, \rangle$ (preselection), (ii) weakly measuring observable $\hat{A}$ of the system, 
(iii) retaining the detector output only if the system is eventually
measured to be in a chosen final state $|\, \textrm{f}\,
\rangle$ (postselection).  
The average signal monitored by the detector will then be proportional
to the real (or imaginary) part of the complex WV,
$\label{eq:weakvalue1} \left.\vphantom{\langle
  \hat{A}\rangle}\right._{\textrm{f}}\langle  \hat{A}
\rangle_{i} = \langle\, \textrm{f} \,| \,\hat{A}\,
| i\, \rangle   /   \langle \,\textrm{f}\, |  i \rangle$,
rather than to the standard average value, $\bra{i} \hat{A} \ket{i}$. Further discussion of the context in which WV should be understood has been provided~\cite{Wiseman:2002,Jozsa:2007,Dressel:2010}.

Going beyond the peculiarities of WV protocols,  recent series of works explored the potential of WVs in quantum optics~\cite{Ritchie:1990,Pryde:2005,Hosten:2008,Dixon:2009,Starling:2009,Brunner:2010,Starling:2010b} and solid-state physics~\cite{Williams:2008,Romito:2008,Shpitalnik:2008,Zilberberg:2011}, ranging from experimental observation to their application to hypersensitive measurements. In the latter, a measurement, performed by a detector \emph{entangled} with a system whose states can be preselected and postselected, leads to an amplified signal in the detector that enables sensing of small quantities~\cite{Hosten:2008,Dixon:2009,Starling:2009,Brunner:2010,Starling:2010b,Zilberberg:2011}.
Quite generally within a WV-amplification protocol, only a subset of the detector's readings, associated with the tail of the signal's distribution, is accounted for. Notwithstanding the scarcity of data points, the large value of ${}_f\langle \hat{A}\rangle_i$, leads to an amplification~\cite{Starling:2009,Zilberberg:2011} of the signal-to-external-noise ratio ($\text{SN}_{\text{ex}}\text{R}$) for systems where the noise is dominated by an external (technical) component.

The amplification originating from WV protocols is non-universal. The specifics of such amplification are diverse and system-dependent.	In fact, for statistical (inherent) noise,  SNR amplification resulting from  large WVs is generally suppressed due to  the reduction in the statistics of the collected data:  postselection restricts us to a small subset of the readings at the detector.  The upside of the WV procedure has several facets: if we try to enhance the statistics  by increasing the intensity of input signal through the system (e.g., intensity of the impinging photon beam), possibly entering a non-linear response regime, postselection will effectively reduce this intensity back to a level accessible to the detector sensitivity~\cite{Dixon:2009}. Alternatively, amplification may originate  from the imaginary component of the weak value~\cite{Hosten:2008},  or from the different effect of the noise and the measured variable on the detector's signal~\cite{Zilberberg:2011}.
 However, as long as quantum fluctuations (leading to inherent statistical noise)
dominate, the large WV is outweighed by the scarcity of data points, failing to amplify the signal-to-statistical-noise ratio~\cite{Zilberberg:2011,Zhu:11}.

We have recently  presented an alternative measurement protocol dubbed \textit{null weak value} (NWV), that leads to high fidelity discrimination between qubit states on the background of \emph{quantum fluctuations}~\cite{Zilberberg:12}. The main features of this protocol are (i) It is distinctly different from the standard WV in making use of a partial-collapse measurement -- a strong measurement that occurs with a small probability realized by finely tuned, time resolved coupling to the detector  --, in which the system experiences  weak backaction only for a subset of all possible measurement outcomes (see e.g.~, Refs.~\cite{Korotkov:2007,Katz08}; (ii) Unlike standard WV-amplification procedures~\cite{Hosten:2008,Dixon:2009,Starling:2009,Brunner:2010,Starling:2010b,Zilberberg:2011}, where one needs to employ two degrees of freedom (related, respectively, to ``system'' and ``detector'') which are entangled by the weak measurement, in NWVs there is a single quantum degree of freedom (``system'') since the detector is classical; (iii) The SNR amplification is versus inherent quantum and statistical fluctuations, and not only against external detector noise.

Here we present a general formalism for NWVs. After briefly reviewing   the derivation of standard-WV (Section~\ref{standard}), in Section~\ref{multi} we generalize the definition of a NWV to a multi-level system.
The main results are summarized in the section~\ref{conclusion}.

\section{Weak values}
\label{standard}

Weak values describe the outcome read in a detector when the measured system is subsequently found to be in a specific state. The weak coupling between a system and a detector is performed by an ideal von Neumann measurement~\cite{von-neumann}, described by the Hamiltonian 
\begin{equation}
\label{hamiltoniana}
H=H_{\textrm{S}}+H_{\textrm{D}}+H_{\textrm{int}} \, , \,\,\,\,\,\,
H_{\textrm{int}}=\lambda\, g(t) \hat{p}\hat{A} \, ,
\end{equation}
where $H_{\textrm{S(D)}}$ is the Hamiltonian of the system (detector), and $H_{\textrm{int}}$ is the interaction Hamiltonian between the two. 
Here $\hat{p}$ is the momentum canonically conjugate to the position of the detector's pointer, $\hat{q}$, and
$\lambda\, g(t)$ ($\lambda\ll1$) is a time dependent coupling constant.  
$\hat{A}=\sum_i a_i \ket{a_i}\bra{a_i}$ is the measured observable.
We assume for simplicity that the free Hamiltonians of the system and the detector vanish and that $g(t)=\delta(t-t_0)$.

The system is initially prepared in the state $\ket{i}$, and the detector in the state $\ket{\phi_0}$. The latter is assumed to be a Gaussian wave-packet centered at $q=q_0$, $\ket{\phi_0}=C e^{-(q-q_0)^2/4\Delta^2}$.
After the interaction with the detector the entangled state
of the two is 
\begin{equation}
\label{stato_entangled}
\ket{\psi}=e^{-i\lambda \hat{p} \hat{A} } \ket{i}\ket{\phi_0} \, .
\end{equation}
In a regular measurement the signal in the detector, i.e. the pointer's position $\langle q \rangle= q_0 +\lambda \langle \hat{A} \rangle$, is read.
From the classical signal, $\langle q \rangle$, one can infer the average value of the observable $\hat{A}$.

In a weak value protocol the signal in the detector is kept provided that the system is successfully postselected to be in a state $\ket{f}$.
Hence, the detector ends up in the state
\begin{align}
\label{spostamento}
\ket{\psi}&=\ket{\phi_0} -i \lambda \left[\langle f|\hat{A}|i \rangle/
\langle f|i \rangle\right] \hat{p}\ket{\phi_0}\nonumber\\
&\approx
e^{-i \lambda {}_f\langle\hat{A}\rangle_i} \hat{p} 
\ket{\phi_0} \, ,
\end{align}
that corresponds to a shift in the position of the pointer proportional to $ \mathop{Re}[_{f} \langle A \rangle_{i} ]$.
Hence the expectation value of the coordinate of the pointer is given by
\begin{equation}
{}_f\langle \hat{q}\rangle_i=q_0+\lambda  \mathop{Re}[_{f} \langle \hat{A} \rangle_{i} ]\, ,
\end{equation}
with
\begin{equation}
{}_f\langle \hat{A}\rangle_i=\frac{\langle f|\hat{A}|i \rangle}{
\langle f|i \rangle}\equiv \text{WV}(\hat{A})\, .
\label{swv}
\end{equation}
Here, too, the conditional average value of $\hat{A}$ is inferred from the detector's reading.

We note that the approximation in Eq.~\ref{spostamento} is valid when $\Delta \gg \max_{i,j} |a_i-a_j|$. 
This means that the initial detector's wave function and the shifted one due to the interaction with the system are strongly overlapping. In turn, this means that for \emph{any} outcome of the detector the state of the system is weakly changed. This corresponds to a weak measurement. As long as the measurement of the observable, $\hat{A}$, is weak, the weak value is a general result and does not depend on the details of the coupling or the specific choice of the detector.


\section{Null weak values in multi-level systems}
\label{multi}

A different measurement protocol based on a postselected readout has been recently proposed as an efficient tool for qubit state discrimination~\cite{Zilberberg:12}.
 The NWV protocol consists, like the standard-WV protocol, of a two-step procedure. Consider first a two-level system. It is initially prepared in a state $\ket{i}$. 
The first measurement, $M_w$, is a strong (projective) measurement which is performed on the system with small probability $p$. Specifically, during the time of this measurement we allow the system to decay, thus inducing a signal in the detector. The detector then ``clicks'' (the measurement outcome is positive) and  the qubit system is destroyed. Very importantly, having a ``null outcome'' (no click) still results in a weak backaction on the system. We refer to this stage of the measurement process as ``weak partial-collapse''.
Subsequently the qubit state is (strongly) measured a second time (postselected), $M_s$, to be in the state $\ket{f}$ (click), $\ket{\bar{f}}$ (no click). The conditional probability of [a ``click'' in the weak partial-collapse measurement conditional to ``no click'' in the postselection], $P(M_{w}|\bar{M}_{s})$, defines the NWV with $M$ and $\bar{M}$ representing ``click'' and ``no click'', respectively. Events in which the qubit is measured strongly (in the second measurement), $M_s$, are discarded.

This protocol substantially differs from the standard-WV in the employment of the weak partial-collapse measurement.
As long as a single measurement is concerned, the partial-collapse measurement will give the same results as a standard von Neumann measurement. However, in the NWV protocol, involving a postselection leads to  results which are qualitatively different as compared with  the application of a standard WV protocol. In fact the NWV protocol allows to discriminate between two possible initial qubit states, $\ket{\psi_0}$ and $\ket{\psi_\delta}$, via repeating the protocol for both cases and comparing the respective conditional outcomes, $P(M_{w,0}|\bar{M}_{s,0})$ and $P(M_{w,\delta}|\bar{M}_{s,\delta})$. 
 Defining the signal $\tilde{S}\equiv P(M_{w,\delta}|\bar{M}_{s,\delta})-P(M_{w,0}|\bar{M}_{s,0})$, it has been shown that the NWV protocol  can be used to discriminate between the two states with an amplified SNR as compared with a standard measurement of a single observable.

\begin{figure}
 \centering
 \includegraphics[width=0.9\columnwidth]{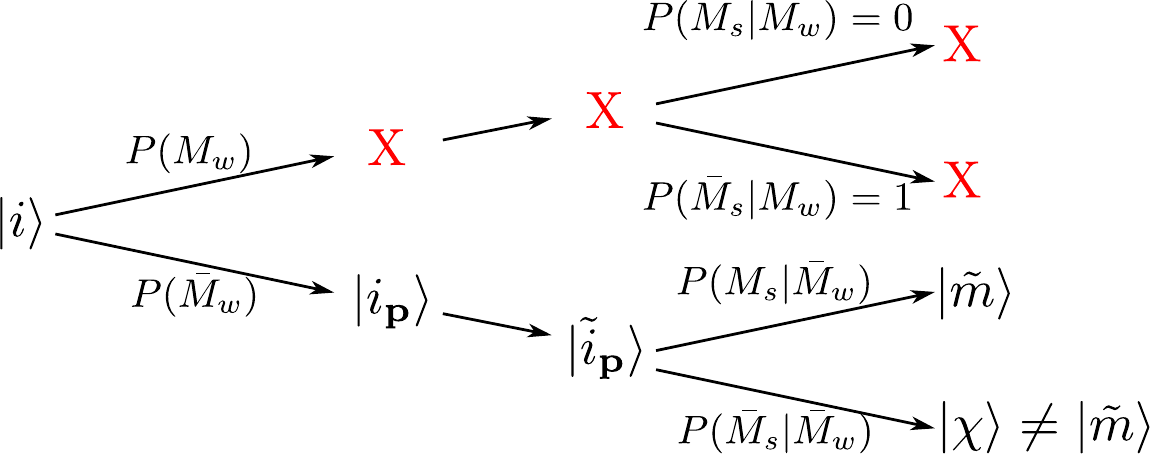}
 \caption[]{\label{Fig:1}
  A tree diagram  of the state evolution under subsequent partial-collapse measurements; the respective  probabilities are indicated: $P(M_{w})$ [$P(\bar{M}_{w})$] is the probability that the detector ``clicks'' [no ``click''] upon the first measurement. If it does ``click'', the system is destroyed, hence there are no clicks upon further measurements [this is marked by a (red) $\mbox{X}$]. Note that following $P(\bar{M}_{w})$ (null detection of the system), the weak backaction rotates $\ket{\psi}$ into $\ket{\psi_\mathbf{p}}$ and then, upon further evolution, into $|\tilde{\psi}_\mathbf{p}\rangle$.}
 \end{figure}

We present here a general formalism for NWVs of a multi-level system.
For the sake of being pedagogical, let us think of a single quantum particle in a closed system. The single-particle states span an $n$-dimensional Hilbert space. The system is initially prepared in a state $\ket{i}=\sum_{m=0}^{n-1}\alpha_m\ket{m}$, $\sum_{m=0}^{n-1}|\alpha_m|^2=1$, with $\braket{m}{m'}=\delta_{m,m'}$. Our protocol is comprised of two consecutive measurements (see Fig.~\ref{Fig:1}). 
The first measurement is a weak partial-collapse measurement. One can think of a partial-collapse measurement as coupling the system to a continuum of states (the ``detector'') through a tunnel barrier. Tunneling is allowed for a time window $t$, effectively enabling the states $\ket{m}$ to tunnel out of the system with probability $p_m=1-\text{Exp}(-\Gamma_m t)$, where $\Gamma_m$ is the corresponding tunneling rate for $m=0,...,n-1$. 
The information on whether the particle is found out of the system can be studied and defines the detector's signal. Hence, the average outcome of the partial-collapse measurement would result in the measurement of the observable
\begin{align}
	\hat{M}_w\equiv\sum_{l=0}^{n-1}p_l\ket{l}\bra{l}\, .
\end{align}
Note that this observable depends on the tunneling probabilities and is therefore not a pure observable of the system.
There are several situations where a calibration of this measurement leads to an observable of the system only. For example, if the tunneling probability of a specific state, $\ket{m}$ is much larger than that of the other states ($p_l \ll p_m$ for any $l\leq m$) we can measure the population of this state $\ket{m}\bra{m}\approx \hat{M}_w /p_m$. Another example of such a calibration is for a case where (without loss of generality) the states $\ket{j}$, $0\leq j \leq k < n-1$, tunnel out with probability $p$ and the other states $\ket{j}$, $k < l \leq n-1$, tunnel out with probability $p'$. Here, the population of the subspace $\sum_{l=0}^{k} \ket{l}\bra{l}=(\hat{M}_w-p') /(p-p')$ is obtained. Having in mind these cases we define, for the sake of brevity, the operator of the system
\begin{align}
	\hat{A}\equiv\frac{\hat{M}_w}{f(\mathbf{p})}\, ,
\end{align}
where $f(\mathbf{p})$ is a calibration function that translates the probabilistic outcome of the partial-collapse procedure (characterized by $\mathbf{p}=p_0,...,p_{n-1}$) to an observable of the system.

In order to include postselection, we assume that subsequently to the tunneling (measurement of the ``system'' by the ``detector'') the system undergoes a general unitary evolution, $\hat{U}(\tau)$. Finally the state of the particle is (strongly) measured a second time (postselected) to be, for example, in state $\ket{\tilde{m}}$, $\hat{M}_s=\ket{\tilde{m}}\bra{\tilde{m}}$. One may now define an expectation value of the operator $\hat{A}$, conditioned on the reading of the second measurement being negative ($\bar{M}_s)$,
\begin{align}
	{}_{\bar{M}_s}\langle \hat{A}\rangle_{i} &\equiv \frac{P\left(M_w |\bar{M}_s\right)}{f(\mathbf{p})}\, ,
	\label{nullwv}
\end{align}
where $P\left(M_w |\bar{M}_s\right)$ is the conditional probability of having a click in the first measurement conditional to \textit{not} having a click the second time. This is the NWV. Note that, in similitude to a standard-WV~\cite{Aharonov:1988aa,Romito:2008}, ${}_{\bar{M}_s}\langle \hat{A}\rangle_{i}$ may give rise to large values, beyond the interval [0,1].

To shed some light on this expression we calculate explicitly the conditional probabilities following the measurement procedure sketched in Fig.~\ref{Fig:1}.
As a first step, we note that when the detector clicks during the first measurement (positive outcome of the detector) the state of the system is destroyed.
Very importantly, having a ``null outcome'' (no click) still results in a weak backaction  (partial collapse) on the system. 
In such a case the state can be written as
\begin{align}
	\ket{i_{\mathbf{p}}}=\frac{1}{\sqrt{P(\bar{M}_w)}}\sum_{m=1}^{n}\alpha_m\sqrt{1-p_m} e^{i\phi_m}\ket{m}\, ,
	\label{modState}
\end{align}
where $P(M_w)=1-P(\bar{M}_w)=\sum_{m=1}^{n}p_m |\alpha_m|^2$ is the probability of the state to be positively measured. We include general relative phases that such a coupling could induce in the form of $\phi_m$.

Rewriting the conditional probability on the right-hand-side of Eq.~\eqref{nullwv} using Bayes theorem, we obtain
\begin{multline}
	P\left(M_w |\bar{M}_s\right)=\\
	\frac{P\left(M_w\right) P\left(\bar{M}_s|M_w\right)}{P\left(M_w\right) P\left(\bar{M}_s|M_w\right)+P\left(\bar{M}_w\right) P\left(\bar{M}_s|\bar{M}_w\right)}\, .
	\label{nullwv2}
\end{multline}
The conditional probability $P(\bar{M}_s|M_w)=1$ represents classical correlation between the two measurements, namely, given that the detector has clicked in the first measurement (the particle has tunneled out to the detector), the result of the second measurement will be negative with probability $1$. By contrast, $P(\bar{M}_s|\bar{M}_w)$ embeds non-trivial quantum correlations.
If no tunneling takes place [with probability $P(\bar{M}_{w,\delta})$] the state evolves into $\ket{i_{\mathbf{p}}}$ of Eq.~\eqref{modState}. During the time $\tau$ between the two measurements, the state is evolved in a controlled fashion to $\ket{\tilde{i}_{\mathbf{p}}}=\hat{U}(\tau)\ket{i_{\mathbf{p}}}$. If the state was measured in the first measurement, this evolution is truncated.
The second strong measurement, $M_s$ gives a positive outcome (``click'' of the detector) with probability $P(M_{s}|\bar{M}_{w})=\left|\braket{\tilde{m}}{\tilde{\psi}_{\mathbf{p}}}\right|^2$ and a negative one with $P(\bar{M}_{s}|\bar{M}_{w})=\sum_{m\neq \tilde{m}}\left|\braket{m}{\tilde{\psi}_{\mathbf{p}}}\right|^2$. Collecting this into Eq.~\eqref{nullwv2}, we obtain
\begin{align}
P(M_{s}|\bar{M}_{w})=\frac{P(M_{w})}{P(M_{w}) + P(\bar{M}_{w})P(\bar{M}_{s}|\bar{M}_{w})}\, ,
\label{nullwv3}
\end{align}
and after calibration the NWV of Eq.~\eqref{nullwv} can be written as
\begin{align}
	{}_{\bar{M}_s}\langle \hat{A}\rangle_{i} &\equiv \frac{\langle\hat{A}\rangle}{P\left(\bar{M}_s\right)}\, .
	\label{nullwv4}
\end{align}

\begin{figure}
 \centering
 \includegraphics[width=\columnwidth]{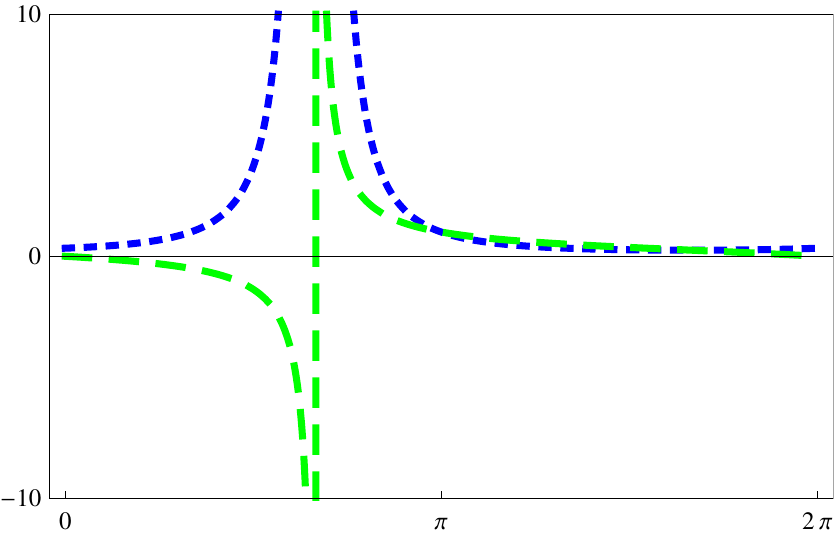}
 \caption[]{\label{Fig:2}
  The real part of the standard-WV (dashed green line) and the NWV (dotted blue line) of the observable $\hat{n}_1$ of a qubit vs.~the postselection angle. The initial state is arbitrarily chosen to be $\ket{i} = \cos(\pi/6)\ket{0}+\sin(\pi/6)\ket{1}$. The postselection angle, determined by the the rotation $U(\tau)$, is defined by $\ket{f} = \cos(\gamma)\ket{0}-\sin(\gamma)\ket{1}$. The protocol is readily generalized to include characterization of the state by both tangential and azimuthal angles.
}
 \end{figure}

An interesting case is that of a two-level systems, where only two tunneling probabilities exist and the NWV protocol uniquely defines the measured observable $\hat{n}_1=\ket{1}\bra{1}=(\hat{M}_w-p_0)/(p_1-p_0)$, i.e.~the population of the state $\ket{1}$.
The postselection is arbitrarily chosen to consist of a projection on the state $\ket{1}$. Taking a weak partial-collapse limit $p_0 , p_1 \ll 1$ such that $P(M_{w}) \ll P(\bar{M}_{w})P(\bar{M}_{s}|\bar{M}_{w})$, Eq.~\eqref{nullwv4} yields
\begin{equation}
	\label{josnullwvapprox}
	{}_{f}\langle \hat{n}_1\rangle_{i} \approx \frac{\bra{i}\hat{n}_1\ket{i}}{|\langle f| i\rangle|^2}\equiv \text{NWV}(\hat{n}_1) \, ,
\end{equation}
 where we have defined the final state to include a general rotation between the partial-collapse measurement and the postselection, $\ket{f}=\hat{U}^\dagger(\tau)\ket{0}$. The expression for the NWV result in Eq.~\eqref{josnullwvapprox} is indeed different from the standard-WV expression [see Eq.~\eqref{swv}]. This is a manifestation of the partial-collapse measurement, to be contrasted with a Von-Neumann type~\cite{von-neumann} weak measurement underlying the latter.
The differences between the two values are shown in Fig~\ref{Fig:2}. It appears  that the NWV diverges with a wider envelope than the standard-WV as the pre- and postselection states become more orthogonal.


\section{Conclusions}
\label{conclusion}
We have presented here a novel measurement protocol, applicable to a general system spanning an $n$-dimensional Hilbert space. Similar to standard weak values, the outcome of this protocol -- null weak value -- is the result of a first (weaker) measurement correlated with a strong postselection. Ostensibly, as long as a single measurement is concerned, the first measurement in both protocols yields the same outcome. However, the substantial difference between the standard- and null- weak values comes to show that its backaction on the system is profoundly different. Hence, involving a postselection leads to qualitatively different correlated results.

\begin{acknowledgments}
This work was supported by GIF, ISF, Minerva Foundation, Israel-Korea MOST grant, and EU GEOMDISS. 
\end{acknowledgments}


%

\end{document}